\documentclass[10pt,letterpaper]{article} 
\usepackage{titlesec,float,makeidx,multirow,multicol,graphicx,hyperref,amsmath,amssymb,fancyhdr,changepage,etoolbox,times,enumitem,textcomp,indentfirst,subcaption,caption,setspace}

\usepackage[dvipsnames,svgnames,table]{xcolor}
\usepackage{txfonts}
\usepackage{mathdots}
\usepackage[paperwidth=8.5in,paperheight=11in,top=1in,right=0.5in,bottom=1in,left=0.5in]{geometry}
\usepackage[utf8]{inputenc}
\usepackage[sort&compress,numbers]{natbib} 
\fancypagestyle{firstpage}{
\fancyhead[L,C,R]{}
\fancyfoot[R]{ }
\fancyfoot[C]{\thepage}
\fancyfoot[L]{}

}

\RequirePackage{fix-cm}
\newcommand{\size}[2]{{\fontsize{#1}{0}\selectfont#2}}

\titleformat{\section}
  {\normalfont\fontfamily{phv}\bfseries}{\thesection}{10pt}{\MakeUppercase}
\renewcommand*{\thesection}{\fontfamily{phv}\selectfont\textbf{\arabic{section}.}}
\titlespacing{\section}{0pt}{10pt}{0pt}
\titleformat{\subsection}
  {\normalfont\fontfamily{phv}\bfseries}{\thesubsection}{3pt}{}
\renewcommand*{\thesubsection}{\fontfamily{phv}\selectfont\textbf{\arabic{section}.\arabic{subsection}}}
\titlespacing{\subsection}{0pt}{10pt}{0pt}

\DeclareCaptionFormat{upper}{#1#2\uppercase{#3}\par}
\hypersetup{
	colorlinks=true,
	linkcolor=black,
	filecolor=blue,      
	urlcolor=blue,
	citecolor=black,
}
\usepackage{nomencl}
\usepackage{bm}
\usepackage{xstring}
\usepackage{xpatch}
\patchcmd{\thenomenclature}
  {\leftmargin\labelwidth}
  {\leftmargin\labelwidth\itemindent 0.25in }
  {}{}

\makenomenclature

\begin{document}
\newgeometry{top=0.5in,right=0.5in,bottom=1in,left=0.5in}
\begin{flushright}
\fontfamily{phv}\selectfont{
\textbf{Conference Paper}\\
\textbf{2022}\\
}
\end{flushright}
\begin{center}
    \fontfamily{phv}\selectfont{\size{11}{\textbf{Dynamic Transition From Mach to Regular Reflection Over a Moving Wedge\\[20pt]}}}
\end{center}


    \begin{flushright}
        \fontfamily{phv}\selectfont{\textbf{Lubna Margha$^{1,2}$, Ahmed A. Hamada$^{1}$, Ahmed Eltaweel$^{3}$} \newline
        $^1$Department of Ocean Engineering, Texas A$\&$M University, College Station, TX, 77843, USA \newline
        $^2$Department of Aeronautical and Aerospace Engineering, Cairo University, Giza 12613, Egypt \newline 
         $^3$Aerospace Engineering Program, University of Science and Technology - Zewail City, Giza, 12578, Egypt \newline }
    \end{flushright}
    
\begin{multicols*}{2}
\section*{Abstract}
The design of supersonic and hypersonic air-breathing vehicles is influenced by the transition between the Mach Reflection (MR) and Regular Reflection (RR) phenomena. The purpose of this study is to investigate the dynamic transition of unsteady supersonic flow from MR to RR over a two-dimensional wedge numerically. The trailing edge of the wedge moves downstream along the $x$-direction with a velocity, $V(t)$ at a free-stream Mach number of $3$. An unsteady compressible inviscid flow solver is used to simulate the phenomenon. Further, the Arbitrary Lagrangian-Eulerian (ALE) technique is applied to deform the mesh during the wedge motion. The dynamic transition from MR to RR is defined by two criteria, the sonic and the Von-Neumann. Moreover, the lag in the dynamic transition from the steady-state condition is studied using various non-dimensional angular velocities, $\kappa$, in the range of [0.1-2]. The lag effect in the shock system is remarkable at the high values of the non-dimensional angular velocity, $\kappa=1.5$ and $2.0$. Furthermore, because the shock is bent upstream during the fast motion of the wedge, the transition from MR to RR happens below the Dual Solution Domain (DSD).

Keywords: Unsteady; Regular reflection; Mach reflection; Moving wedge; Dynamic shock waves; Supersonic flow; Dual solution domain.
\section{Introduction}
\begin{figure}[H]
	\centering
	\begin{subfigure}{\linewidth}
		\centering
		\includegraphics[width=\linewidth]{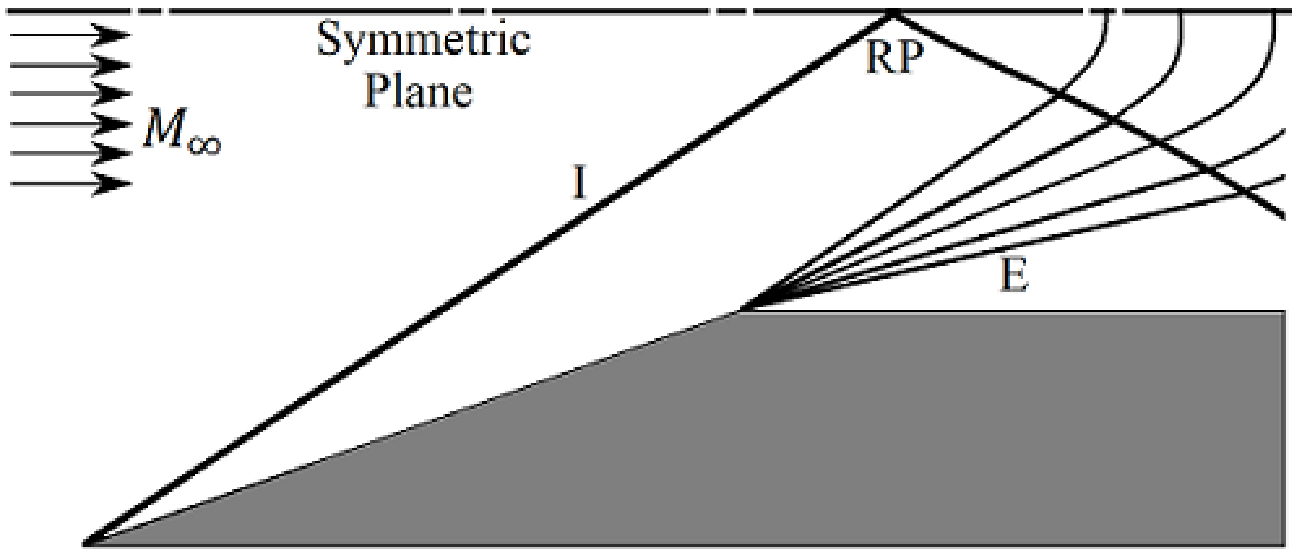}
		\caption{ RR shock structure.}\label{fig:Fig1a}
	\end{subfigure}\\
	\begin{subfigure}{\linewidth}
		\centering
		\includegraphics[width=0.95\linewidth]{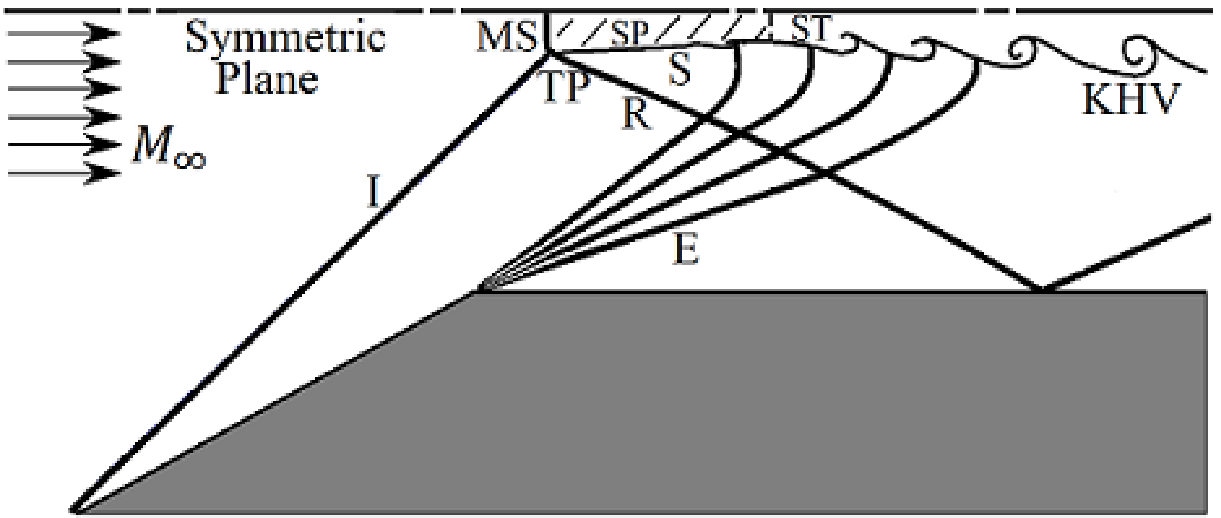}
		\caption{ MR shock structure.}\label{fig:Fig1b}
	\end{subfigure}
	\caption{Schematic of the supersonic flow field over a wedge showing the RR and MR shock structures.}\label{fig:Fig1}
\end{figure}

The accurate predictions of the Regular and Mach reflections transition phenomena are indispensable in many engineering applications, such as supersonic and hypersonic vehicles, the explosion gas dynamics, and shock wave focusing. When a supersonic flow impinges a symmetric sharp wedge of a fixed small deflection angle, ($\theta$), an incident straight oblique shock wave generates and reflects at the top symmetric plane, forming a regular reflected shock wave (RR case, see Figure \ref{fig:Fig1} (a)) for weak shock waves. While, a Mach stem height is initiated when the incident shock wave deflects by an enough large fixed wedge angle, generating a three shock wave configuration and slip-line at the triple point, which is known as the  Mach reflection shock wave (MR case, see Figure \ref{fig:Fig1} (b)). The structure of shock reflections over a symmetric wedge is sensitive to the incoming shock wave's Mach numbers ($M_\infty$), compression ramp angles ($\theta$), and initial boundary conditions. Several scholars clearly investigated the hysteresis of the transition between RR and MR over a stationary wedge, such as Ben-Dor \cite{ben1999hysteresis, ben2002hysteresis, ben2007shock}, Ivanov et al. \cite{ivanov2001transition, ivanov1997transition}, and Yan et al. \cite{yan2003effect, yan2003control}. 

The unsteady dynamic transition from MR and RR is conducted with different types of wedge motion. Naidoo and Skews \cite{naidoo2011dynamic, naidoo2014dynamic} studied numerically and experimentally the unsteady supersonic flow over an impulsive rotating symmetric wedge at different pivot points locations, such as leading edge and trailing edge. They rapidly rotate the leading/trailing edge point at different rates and discussed the effect of choosing the rotating rate on the unsteady flow features and the dynamics of the shock system. Their results concluded that the transitional point and the Mach stem height highly depend on the rotation speed. Moreover, the rotating pivot location affects the development of the flow field. The transition wave angle from MR to RR was delayed and happened below the theoretical von-Neumann limit in the Dual Solution Domain (DSD). Additionally, their results mentioned that the MR configuration can exist for a while at a zero wedge deflection angle at a sufficiently large rotation speed.
\end{multicols*}
\restoregeometry
\clearpage
\begin{multicols*}{2}
Furthermore, a new mechanism was proposed by Margha et. al. \cite{margha2021dynamic} to control the transition between the RR and MR by changing the wedge deflection angle,$\theta$ at constant wedge height, $h$. They numerically studied the RR to MR transition by fixing the wedge leading edge and moving the trailing edge point upstream with a velocity, $V (t)$, and various frequencies, $\kappa$. Their results showed the lag effects of using $\kappa$ on the transition flow parameters. To extend our work, this paper is written to investigate in detail the dynamic effects of $\kappa$ on the transition from MR to RR. A two-dimensional symmetric wedge with an initial inclination angle of $\theta=23^{\circ}$ is exposed to a free-stream Mach number, $M_{\infty}=3$, and the trailing edge point is moved horizontally downstream with a range of $\kappa=$ [0.1-2].

\section{Computational Model}
Two-dimensional unsteady Euler equations were used to set the initial steady-state solution at a wedge angle of $\theta=23^{\circ}$. Then, the horizontal downstream wedge motion was modeled with different values of non-dimensional angular velocities, $\kappa$. The motion was added to the solver by applying the Arbitrary Lagrangian-Eulerian (ALE) technique to compute the new wedge location at each time step. Details of the numerical method, its verification, and the grid generation were provided in the previous work \cite{margha2021dynamic}. 
\subsection{Model Description}
\begin{figure}[H]
    \centering
    \includegraphics[width=0.99\linewidth]{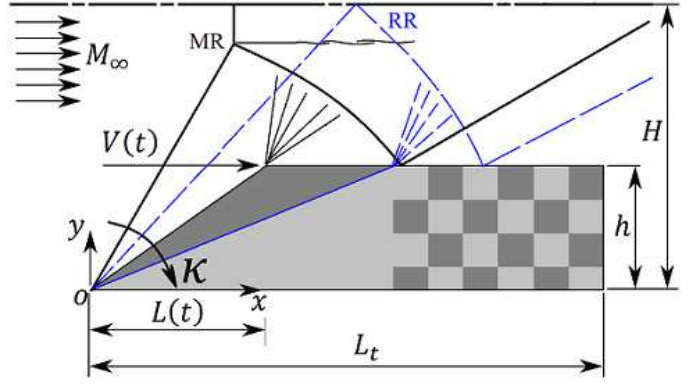}
    \caption{ The flow configuration of the horizontal motion of two symmetrical wedges in a supersonic flow.}\label{fig:fig2}
\end{figure}

The wedge's motion and the two possible flow structures are described in Figure \ref{fig:fig2}. A wedge of fixed height, $h$, and varying length and wedge's angle with time, $L(t)$ and $\theta(t)$, respectively are exposed to a supersonic flow with a Mach number of $3$. The half-height of the computational domain is, $H$, and the total length of both the wedge and the following flat plate is constant during the motion as $L_t$. The motion is started with a steady-state wedge of $\theta_i=23^\circ$ with the Mach Reflection MR shock configuration. Then, the trailing edge is suddenly moved downstream, decreasing the wedge angle, with velocity, $V (t)$, and at different constant rates, $\kappa = [0.1,0.5,1.0,1.5,2.0]$. During the motion $h$ and $L_t$ are kept fixed, while $L(t)$ is increasing and $\theta(t)$ is decreasing with time. All important parameters are given in Table \ref{table:1}.

\begin{table}[H]
\centering
\caption{System properties and parameters.}
\begin{tabular}{c c} 
 \hline \rule{0mm}{2.5ex}
 Initial wedge's chord, $w(0)$                 & $0.833 m$             \\[0.5ex]
 \hline
 Initial wedge angle, $\theta(0)$              & $23^{\circ}$      \\[0.5ex]
 Final wedge angle, $\theta(t_f)$              & $10.5^{\circ}$      \\[0.5ex]
 Wedge height to half domain height, $\frac{h}{H}$ & $0.3617$      \\[0.5ex]
 Initial wedge length to half domain height, $\frac{L(0)}{H}$ & $0.85221$\\[0.5ex]
 Total wedge length to half domain height, $\frac{L_t}{H}$ & $2$   \\[0.5ex]
 Free-stream Mach number, $M_{\infty}$         & $3$               \\[0.5ex]
 Reduce frequency, $\kappa$                    & $\left\{\begin{matrix} 0.1,0.5,\\1.0,1.5,\\2.0\end{matrix}\right\}$       \\[0.5ex]
 \hline
\end{tabular}
\label{table:1}
\end{table}
\subsection{Governing Equations}
Two-dimensional unsteady Euler equations for supersonic flows are used to compute the flow over the wedge and are expressed as:	
\begin{equation}
	\frac{\partial Q}{\partial t}+\frac{\partial F}{\partial x}+\frac{\partial G}{\partial y}=0,
\end{equation}
where
\begin{equation}
Q= \begin{bmatrix}
    \rho \\
    \rho u\\
    \rho v\\
    \rho e
\end{bmatrix}, \quad F= \begin{bmatrix}
    \rho u \\
    \rho u^2+p\\
    \rho u v\\
    u (\rho e+p)
    \end{bmatrix}, \quad G= \begin{bmatrix}
    \rho v \\
    \rho u v\\
    \rho v^2+p\\
    v (\rho e+p)
    \end{bmatrix}
\end{equation}
The static pressure is obtained from
\begin{equation}
	p=(\gamma-1) \left( \rho e - \rho \frac{u^2+v^2}{2}\right)
\end{equation}
where $p$, $\rho$, and $e$ are the flow field pressure, density, and internal energy, respectively. $u$ and $v$ are the velocity components in the Cartesian coordinates $x$ and $y$, respectively, and $\gamma$ is the gas-specific heat ratio, which is set for a perfect gas of $1.4$.

\subsection{Equations of Motion}
The wedge's trailing edge point is moved horizontally downstream with velocity $V(t)$ and a constant wedge angular velocity, $\omega=d\theta/dt$ ($sec^{-1}$). Accordingly, the velocity of the wedge's trailing edge is expressed as:

\begin{equation}
	V_t(t)=\omega\ h \sqrt{1+cot^{2}(\theta(t))}
\end{equation}
where $h$ is the height of the wedge and it is kept fixed, and $\theta(t)$ is the decreasing wedge angle.\\ 

Further, $\kappa$ is the non-dimensional angular velocity which is normalized using the free-stream velocity, $V_\infty$, and the initial wedge stream-wise length,$L(0)$.

\begin{equation}
	\kappa=\frac{\omega\ L(0)}{U_\infty}
\end{equation} 

Additionally, the time dependant wedge angle as a function of non-dimensional time, $\tau$., is defined as: 
\begin{equation}
	\theta(\tau) = \theta(0) + \kappa \tau
\end{equation}

where $\theta(0)$ is the initial wedge angle at $\tau=0$, and the non-dimensional time is defined as:
\begin{equation}
	\tau=\frac{t\ U_\infty}{L(0)} 
\end{equation}
\subsection{Computational Domain}
\begin{figure}[H]
    \centering
    \includegraphics[width=0.99\linewidth]{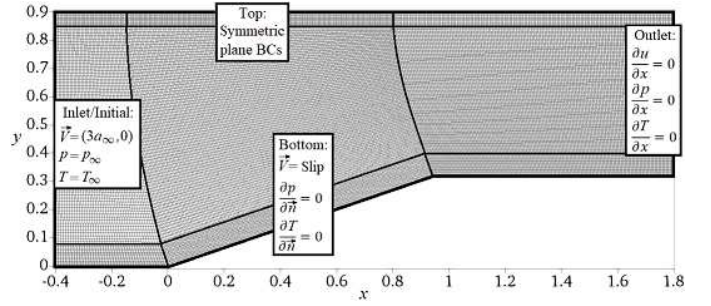}
    \caption{Schematic of the computational domain, the boundary, and initial conditions, for mesh 1}\label{fig:fig3}
\end{figure}

The unsteady supersonic flow with a free-stream Mach number of $3$ over a moving wedge was simulated using the \textit{rhoCentralDyMFoam} solver. It is a density-based solver in the free open-source CFD toolbox, OpenFOAM\textsuperscript{\textregistered}$-$v2006. Kurganov and Tadmor's \cite{kurganov2000new, kurganov2001semidiscrete, greenshields2010implementation} semi-discrete and upwind-central non-staggered techniques were implemented in  the solver. A dynamic mesh was conducted during the computational time using the Arbitrary Lagrangian-Eulerian (ALE) technique \cite{hirt1974arbitrary}, due to using the "\textit{DyM}" solver. Due to the obvious symmetry of the flow behavior and geometry, half of the computational domain was considered. Nine structured and curved blocks with quadrilateral cells were used to discretize the domain and to improve the orthogonality of the cells, as shown in Figure \ref{fig:fig3}. Moreover, this figure shows the boundary conditions over the faces of the domain. The mesh was refined to the level that the absolute percentage error of the Mach stem height, MS, reached $3.4\%$. This is an accepted margin of error, especially since the percentage of error in the tangent wave angle is $0.17\%$, as discussed in detail in our previous work \cite{margha2021dynamic}. The number of cells of the selected grid is $2624 \times 720$ cells. Further details about the solver, the mesh generation, and the verification were discussed by Margha et al. \cite{margha2021dynamic}. 

\section{Results and Discussion}
The dynamic transition from Mach to Regular reflection was investigated using a horizontal downstream moving wedge at a free-stream Mach number of $M_\infty=3$. The problem was simulated by starting with MR in a steady flow at $\theta=23^{\circ}$, then the deflection wedge angle was suddenly reduced to $10.5^{\circ}$. The transition from MR to RR was determined using both the sonic and Von-Neumann criteria. The lag effect in the transition angles, $\theta_t$ and $\beta_t$, and the Mach stem height, MS, was studied using different non-dimensional angular velocities, $\kappa=[0.1,0.5,1.0,1.5,2.0]$.
\subsection{Dynamic Transition from RR to MR}
\begin{figure}[H]
    \centering
    \includegraphics[width=0.9\linewidth]{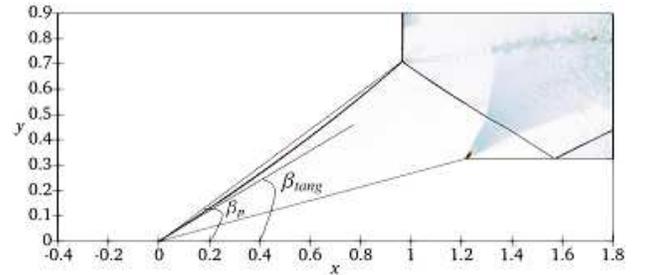}
    \caption{The pressure gradient at $\kappa=2.0$ and $\theta=15^{\circ}$, to show the curvature in the incident wave and the difference between $\beta_p$ and $\beta_{\MakeLowercase{t_{tang}}} $. }\label{fig:fig3b}
\end{figure}

The impulsive motion of the trailing edge point was started from the steady state at $\theta=23^{\circ}$, where the MR configuration exists. The tested non-dimensional angular velocities, $\kappa$ affect the unsteady shock wave configuration causing an obvious lag from the steady-state values. This lag resulted in curvature in the incident shock wave angle as shown in Figure \ref{fig:fig3b}. That's why the tangent wave angle, $\beta_{tang}$, was also measured during the unsteady simulations and was compared with the straight wave angle, $\beta_{p}$ at the same $\theta$ and $\kappa$.    

\end{multicols*}

\begin{figure}[H]
	\centering
	\begin{subfigure}{0.33\linewidth}
	    \centering
        \includegraphics[width=0.99\linewidth]{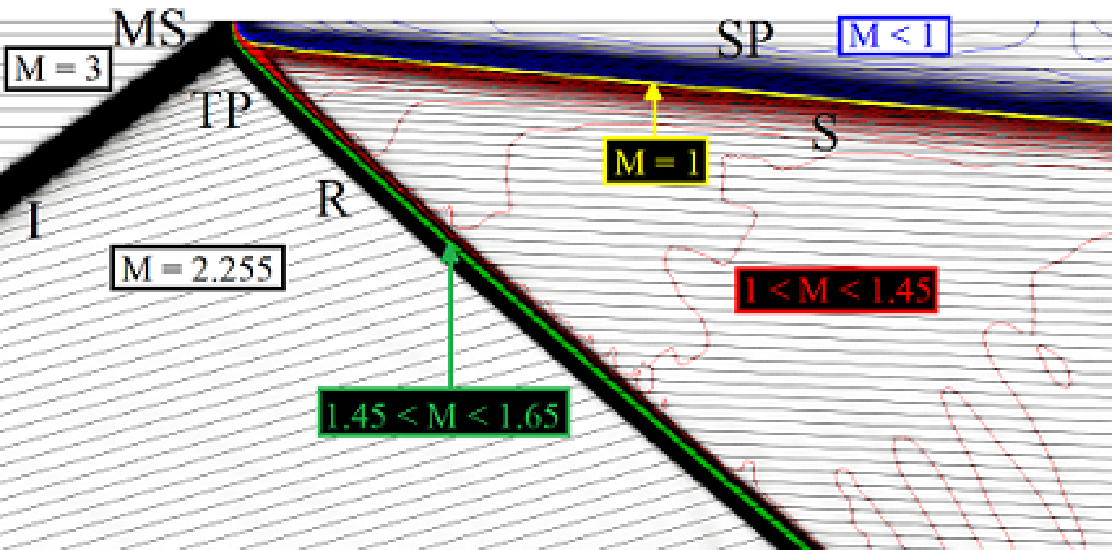}
        \caption{$\theta=15.0014^{\circ}$, MR Case}\label{fig:fig4a}
	\end{subfigure}
	\begin{subfigure}{0.33\linewidth}
	    \centering
        \includegraphics[width=0.99\linewidth]{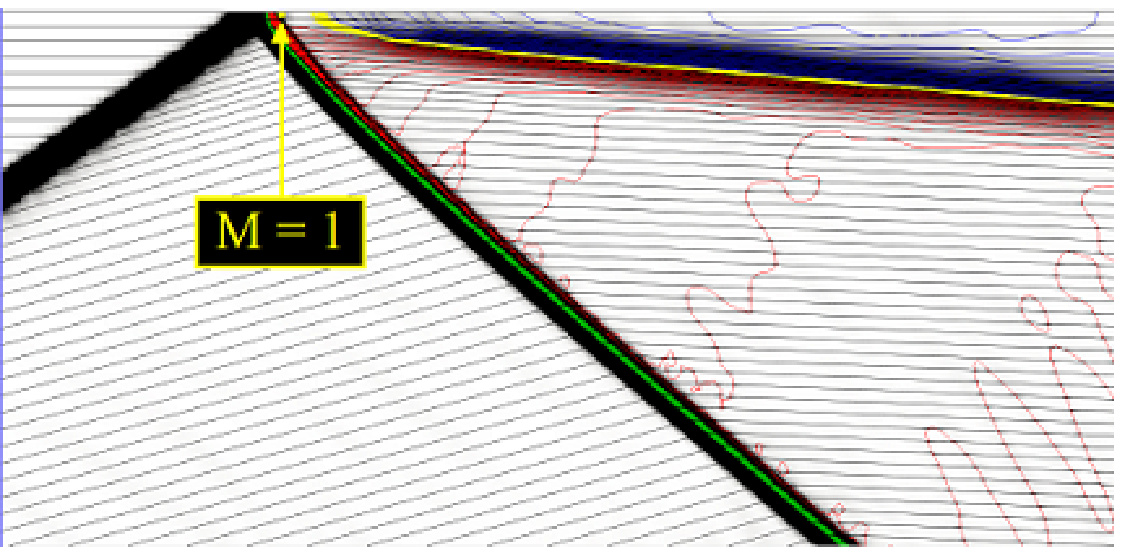}
        \caption{$\theta=14.9878^{\circ}$, Sonic Transition}\label{fig:4b}
	\end{subfigure}
	\begin{subfigure}{0.33\linewidth}
    	\centering  
        \includegraphics[width=0.99\linewidth]{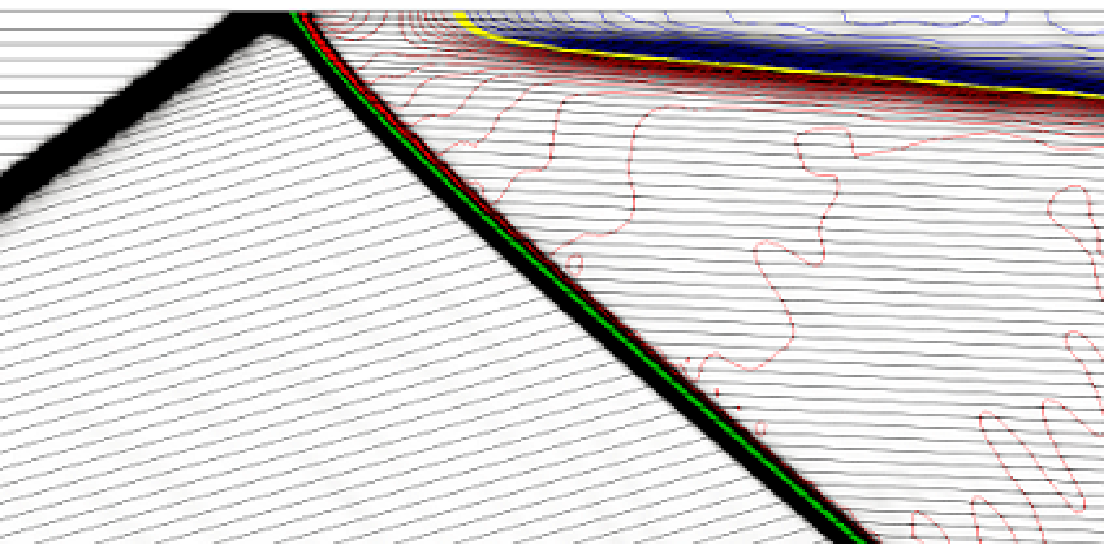}
        \caption{$\theta=14.9743^{\circ}$}\label{fig:4c}
	\end{subfigure}\\
	\begin{subfigure}{0.33\linewidth}
	    \centering
        \includegraphics[width=0.99\linewidth]{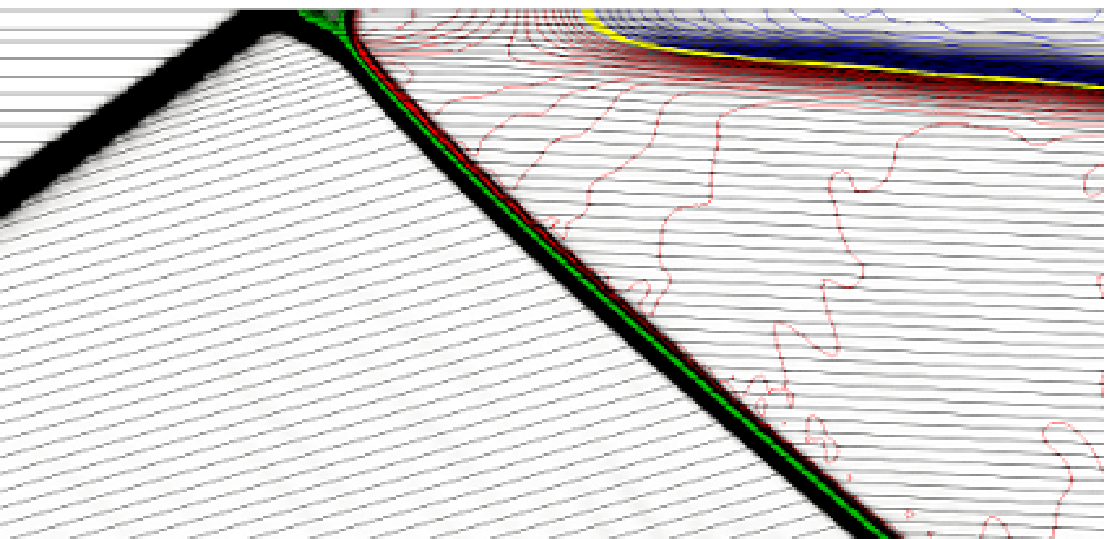}
        \caption{$\theta=14.9607^{\circ}$, von-Neumann Transition}\label{fig:4d}
	\end{subfigure}
	\begin{subfigure}{0.33\linewidth}
	    \centering
        \includegraphics[width=0.99\linewidth]{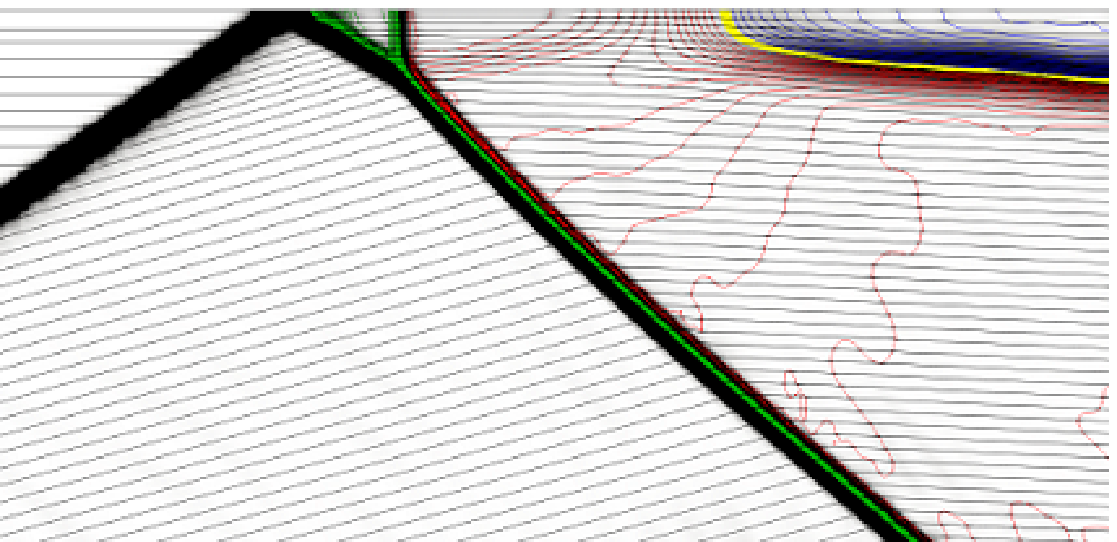}
        \caption{$\theta=14.9471^{\circ}$, RR Case}\label{fig:4e}
	\end{subfigure}
	\begin{subfigure}{0.33\linewidth}
	    \centering
        \includegraphics[width=0.99\linewidth]{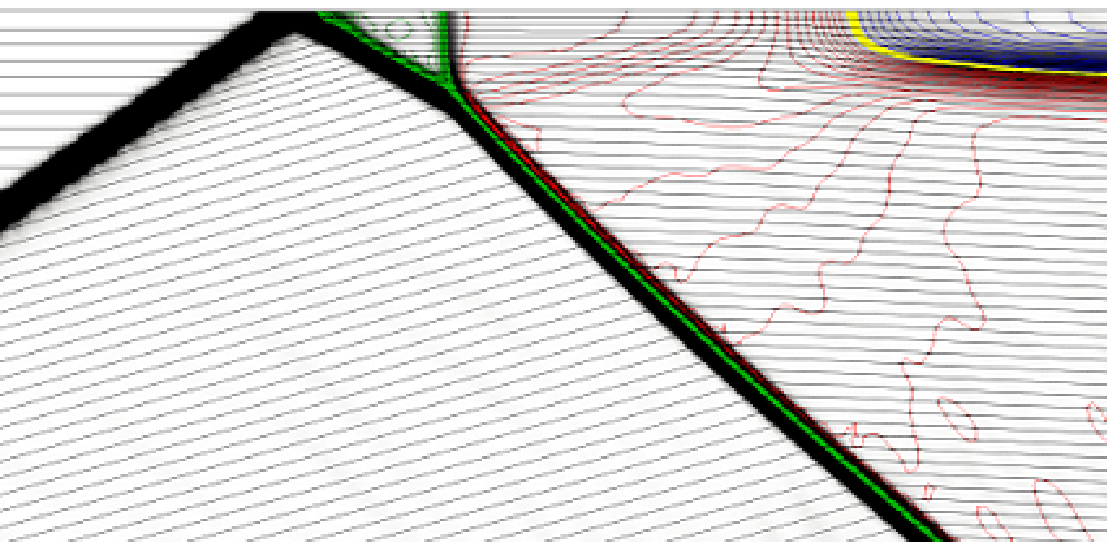}
        \caption{$\theta=14.9335^{\circ}$, RR Case}\label{fig:4f}
	\end{subfigure}
	\caption{Mach contours when the dynamic transition from MR to RR occurs with $\kappa = 0.5$ at Sonic and von-Neumann criteria (Zoomed-in View). }\label{fig:fig4}
\end{figure}

\begin{multicols*}{2}
The Sonic and von-Neumann criteria were used to define the transition from MR to RR. The Sonic condition was measured when the flow behind the reflected shock becomes sonic, $M_{\infty}=1$. While the von-Neumann condition occurred at the point when the slip-line starts to disappear and the flow beyond the Mach stem turns parallel to the mid-plane of symmetry. In the steady state, the von-Neumann condition is known as the physical limit for the steady Mach reflection. Figure \ref{fig:fig4} shows a close view of the used two criteria in measuring the transition from MR to RR at $\kappa=0.5$. The Sonic and von-Neumann limits happened at wedge angles $\theta_{t_{S}}=14.9878$ and $\theta_{t_{vN}}=14.9607$, as shown in Figures \ref{fig:fig4} (b) and \ref{fig:fig4} (d), respectively. 

The lag in the transition wedge and wave angles was summarized in Table \ref{table:2} using the Sonic transition criteria and in Table \ref{table:3} using the von-Neumann transition criteria. The results show that the difference between the dynamic transition angles using the two criteria was within a degree. The steady-state von-Neumann transition for $M_\infty=3$ occurs at a wedge angle of $\theta_{t_{SC}}=19.66^{\circ}$, and an incident wave angle of $\beta_{t_{SC}}=39.34^{\circ}$. Decreasing the wedge deflection angle suddenly with different non-dimensional angular velocities, $\kappa$ resulted in a deviation of these angles from the von-Neumann steady-state transition angles. This deviation increased with the increase in the value of $\kappa$, as indicated in Tables \ref{table:2} and \ref{table:3}. Furthermore, moving the trailing edge of the wedge with a small value of $\kappa=0.1$ caused a deviation in the transition wedge angle, $\theta_t-\theta_{t_{SC}}$, within a degree and in the transition wave angle, $\beta_t-\beta_{t_{SC}}$ of $2.73^{\circ}$. For a relatively higher $\kappa=1.0$, the lag increased, reaching around $8^{\circ}$ in the wedge angle and within $8.9^{\circ}$ in the wave angle. At $\kappa > 1.0$, such as the tested values of $\kappa=(1.5,\ 2.0)$, the transition happens at a deflection wedge angle $\theta < 10.5^{\circ}$. This can not be studied with the current geometries, as the possible minimum wedge angle with the fixed wedge height, $h$, is $10.25^{\circ}$ as shown in Figure \ref{fig:fig10}, using the velocity gradient contours for $\kappa=2.0$. As $\theta$ decreased at a low value of $\kappa=0.1$, the MS moved downstream decreasing its height until vanished and transition to RR occurred as shown in Figure \ref{fig:fig11}. Further, Figure \ref{fig:fig12} shows the transition at a relatively high value of $\kappa=1.0$, while decreasing the wedge incident angle, the MS height increased then abruptly decreased till the transition happened at the end of the motion, $\theta_t=10.7^{\circ}$. 

\begin{table}[H]
\centering
\caption{Sonic transition angles at different values of $\kappa$.}
\begin{tabular}{c c c c c } 
 \hline \rule{0mm}{2.5ex}
 $\kappa$ & $\theta_{\MakeLowercase{t}}$ & $\beta_{\MakeLowercase{t_{p}}}$ &   $\beta_{\MakeLowercase{t_{tang}}} $ & $\tau_t ($ms$)$       \\[0.5ex]
 \hline
    $0.1$  & $18.7794^{\circ}$ &   $36.6112^{\circ}$ & $36.436^{\circ}$  & $42.2064$  \\[0.5ex]
    $0.5$  & $14.9878^{\circ}$ &   $33.4599^{\circ}$ & $31.8521^{\circ}$  & $16.0243$    \\[0.5ex]
    $1.0$  & $10.7645^{\circ}$  &   $30.465^{\circ}$ & $24.6111^{\circ}$  & $12.2355$     \\[0.5ex]
  
 \hline
\end{tabular}
\label{table:2}
\end{table}

\begin{table}[H]
\centering
\caption{von-Neumann transition angles at different values of $\kappa$ .}
\begin{tabular}{c c c c c } 
 \hline \rule{0mm}{2.5ex}
 $\kappa$ & $\theta_{\MakeLowercase{t}}$ & $\beta_{\MakeLowercase{t_{p}}}$ &   $\beta_{\MakeLowercase{t_{tang}}} $ & $\tau_t$       \\[0.5ex]
 \hline
    $0.1$  & $18.7698^{\circ}$ &   $36.6171^{\circ}$ & $36.4229^{\circ}$  & $42.3014$  \\[0.5ex]
    $0.5$  & $14.9607^{\circ}$ &   $33.4338^{\circ}$ & $31.8179^{\circ}$  & $16.0786$    \\[0.5ex]
    $1.0$  & $10.683^{\circ}$  &   $30.3767^{\circ}$ & $24.511^{\circ}$  & $12.317$     \\[0.5ex]
  
 \hline
\end{tabular}
\label{table:3}
\end{table}

\begin{figure}[H]
    \centering
    \includegraphics[width=0.97\linewidth]{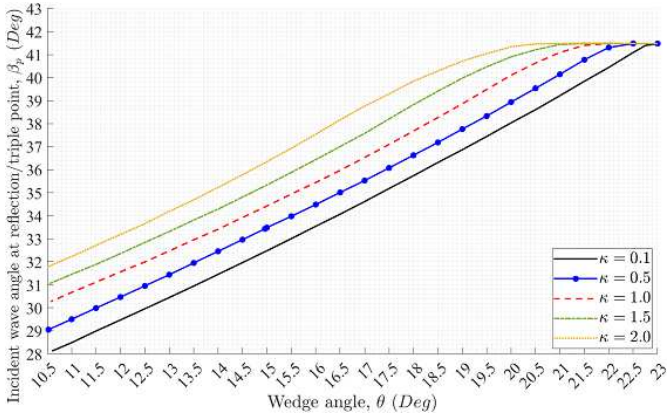}
    \caption{The variation of the incident wave angle at reflection/triple point, $\beta_p$ with decreasing wedge angle, $\theta$, using different values of $\kappa$. }\label{fig:fig6}
\end{figure}

\begin{figure}[H]
    \centering
    \includegraphics[width=0.9\linewidth]{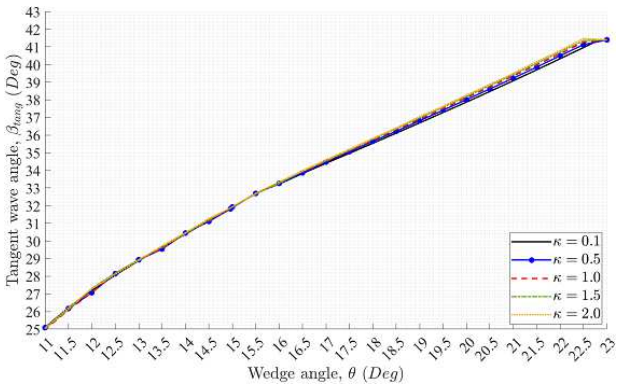}
    \caption{The variation of the tangent incident wave angle, $\beta_{tang}$ with decreasing wedge angle, $\theta$, and different values of $\kappa$. }\label{fig:fig8}
\end{figure}

\begin{figure}[H]
    \centering
    \includegraphics[width=0.99\linewidth]{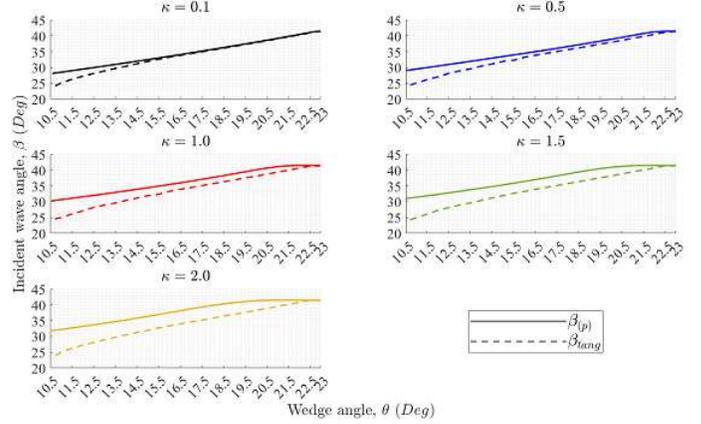}
    \caption{Comparison of the incident wave angle, $\beta$ measured at the reflection/triple point and at the tangent of the incident wave shock, with different values of  $\kappa$.}\label{fig:fig7}
\end{figure}

In addition, the temporal variation of the wave angle, $\beta_p$ during the wedge's trailing-edge motion for different values of $\kappa$ is shown in Figure \ref{fig:fig6}. It showed the lag in the dynamic shock system with wedge motion. This lag affected the transfer of information from the wedge apex to the reflected/triple point. The results indicated that there was no variation in the wave angle, $\beta_p=41.478$, at the beginning period of motion, from $\theta=23^{\circ}$ to a couple of lower angles. This is because the lag was developing during the period due to the sudden motion of the wedge with certain $\kappa$. For example, at high value of $\kappa=2.0$, $\beta_p$ stayed constant at $41.4^{\circ}$ during decreasing the wedge angle from $23^{\circ}$ to $20^{\circ}$. The effect of the lag decreased using lower values of $\kappa$. Moreover, Figure \ref{fig:fig7} assures the concept of the lag during the wedge's motion, by comparing the straight and tangent wave angles at the reflection/triple point, $\beta_p$ and $\beta_{tang}$, respectively. The deviation between the two wave angles grew with the increase of $\kappa$ from within $4^{\circ}$ in the case of $\kappa=0.1$ to within $9^{\circ}$ in the case of $\kappa=2.0$ at the final time of the motion of $\theta = 10.5^{\circ}$. Further, Figure \ref{fig:fig8} showed that the variation in the value of $\kappa$ insignificantly affected the lagged tangent wave angle as the curves were almost the same.

\subsection{Mach Stem Height}
\begin{figure}[H]
    \centering
    \includegraphics[width=0.9\linewidth]{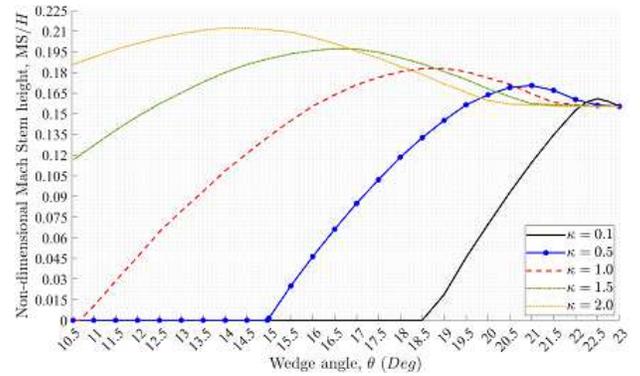}
    \caption{The variation of non-Dimensional Mach stem height with wedge angle, with different non-dimensional angular velocities, $\kappa$.} \label{fig:fig5}
\end{figure}

The dynamic effect of the downstream impulsive motion of the trailing edge on the development of the non-dimensional Mach Stem height, MS$/ H$ is shown in Figure \ref{fig:fig5}. At a small value of $\kappa=0.1$, the MS invisibly increased by $3.5\%$ within a degree, then it rapidly decreased till vanished at the transitional wedge angle of $\theta_t=18.78^{\circ}$. The same manner happened at a higher $\kappa=0.5$, the MS ascended with $9.7\%$ within $2^{\circ}$ of the decreasing wedge angle and abruptly decreased till vanishing and RR happens. Thus, increasing the impulsive motion speed, $\kappa$, results in decreasing the dynamic transitional wedge angle, where the MS disappears and RR occurs. Moreover, the lag strength was significant using high values of $\kappa=(1.5,\ 2.0)$, as the MS remained constant for a while before the obvious growth of the MS height during the descending of the wedge incidence. Particularly, For $\kappa=2.0$, The MS value at the minimum wedge angle of $\theta=10.5^{\circ}$ was $19.3\%$ higher than the initial value of the MS at the steady-state wedge angle of $\theta_i=23^{\circ}$.   
\subsection{Shock Reflection Domain}
\begin{figure}[H]
    \centering
    \includegraphics[width=0.9\linewidth]{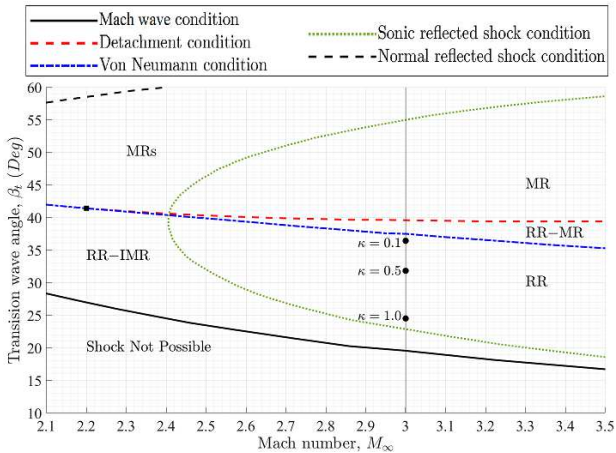}
    \caption{Comparison of dynamic transition wave angle, $\beta_{t_{tang}}$ with the theoretical steady transition wave angle, $\beta_{t_{sc}}$.}\label{fig:fig9}
\end{figure}

\begin{figure}[H]
	\centering
	\begin{subfigure}{0.27\linewidth}
	    \centering
        \includegraphics[width=0.99\linewidth]{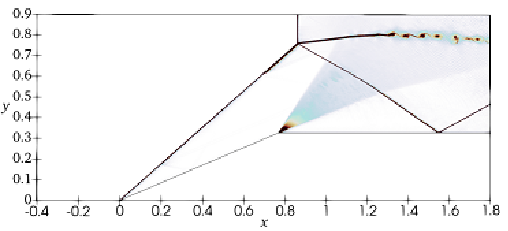}
        \caption{$\theta=23^{\circ}$}\label{fig:fig11_23}
	\end{subfigure}
	\begin{subfigure}{0.27\linewidth}
	    \centering
        \includegraphics[width=0.99\linewidth]{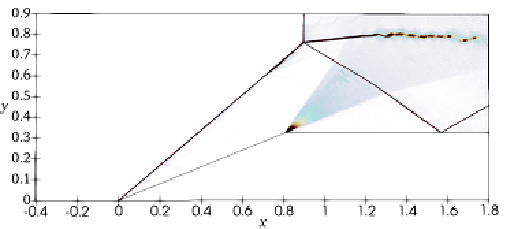}
        \caption{$\theta=22^{\circ}$}\label{fig:fig11_22}
	\end{subfigure}
	\begin{subfigure}{0.27\linewidth}
    	\centering  
        \includegraphics[width=0.99\linewidth]{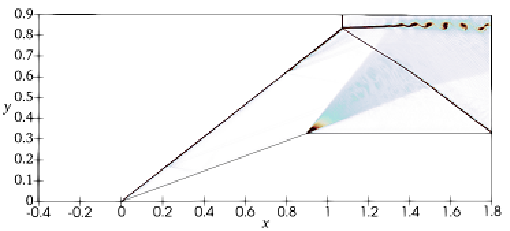}
        \caption{$\theta=20^{\circ}$}\label{fig:fig11_20}
	\end{subfigure}\\
	\begin{subfigure}{0.27\linewidth}
	    \centering
        \includegraphics[width=0.99\linewidth]{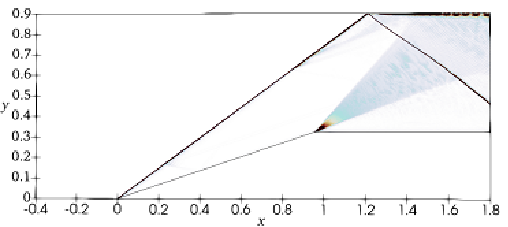}
        \caption{$\theta_t=18.7698^{\circ}$ }\label{fig:fig11_tVN}
	\end{subfigure}
	\begin{subfigure}{0.27\linewidth}
	    \centering
        \includegraphics[width=0.99\linewidth]{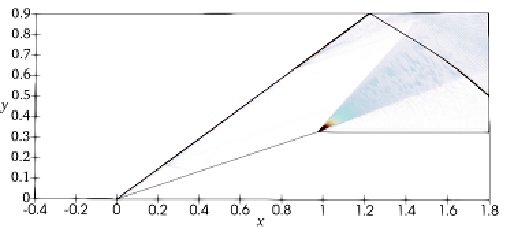}
        \caption{$\theta=18.5^{\circ}$}\label{fig:fig11_18_5}
	\end{subfigure}
	\begin{subfigure}{0.27\linewidth}
	    \centering
        \includegraphics[width=0.99\linewidth]{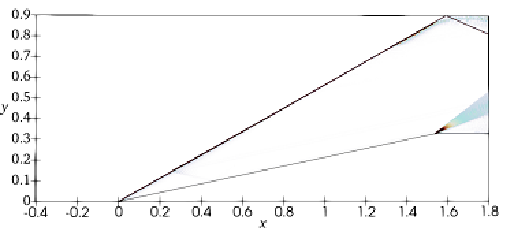}
        \caption{$\theta=12^{\circ}$}\label{fig:fig11_12}
	\end{subfigure}
	\caption{Velocity gradient contours of the dynamic shock system at a low non-dimensional angular velocity, $\kappa = 0.1$, showing the dynamic transition from MR to RR. }\label{fig:fig11}
\end{figure}

\begin{figure}[H]
	\centering
	\begin{subfigure}{0.27\linewidth}
	    \centering
        \includegraphics[width=0.99\linewidth]{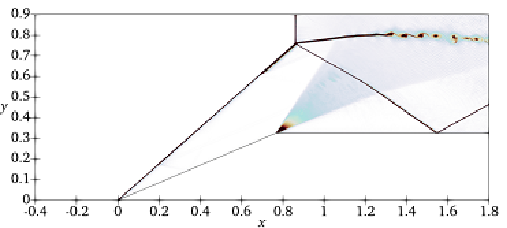}
        \caption{$\theta=23^{\circ}$}\label{fig:fig12_23}
	\end{subfigure}
	\begin{subfigure}{0.27\linewidth}
	    \centering
        \includegraphics[width=0.99\linewidth]{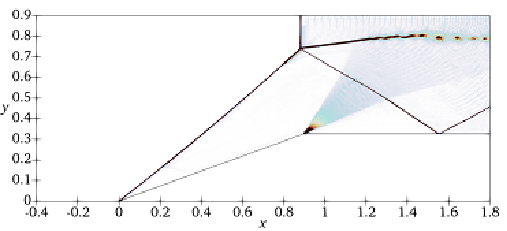}
        \caption{$\theta=20^{\circ}$}\label{fig:fig12_20}
	\end{subfigure}
	\begin{subfigure}{0.27\linewidth}
    	\centering  
        \includegraphics[width=0.99\linewidth]{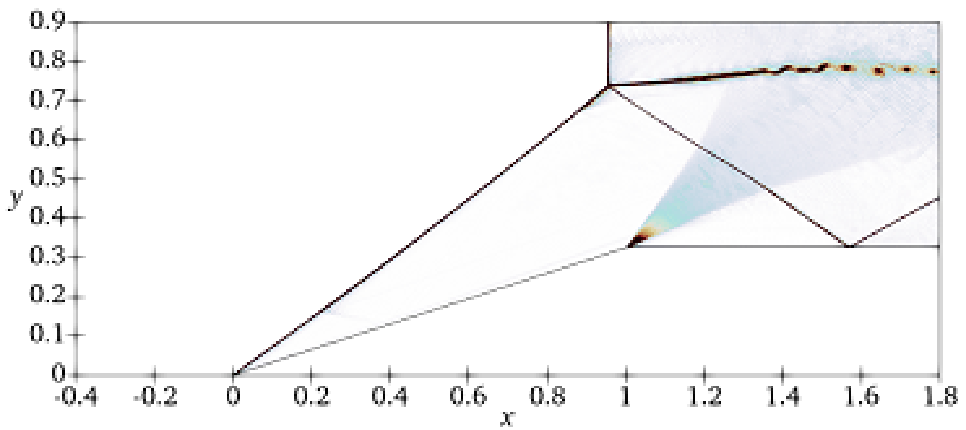}
        \caption{$\theta=18^{\circ}$}\label{fig:fig12_18}
	\end{subfigure}\\
	\begin{subfigure}{0.27\linewidth}
	    \centering
        \includegraphics[width=0.99\linewidth]{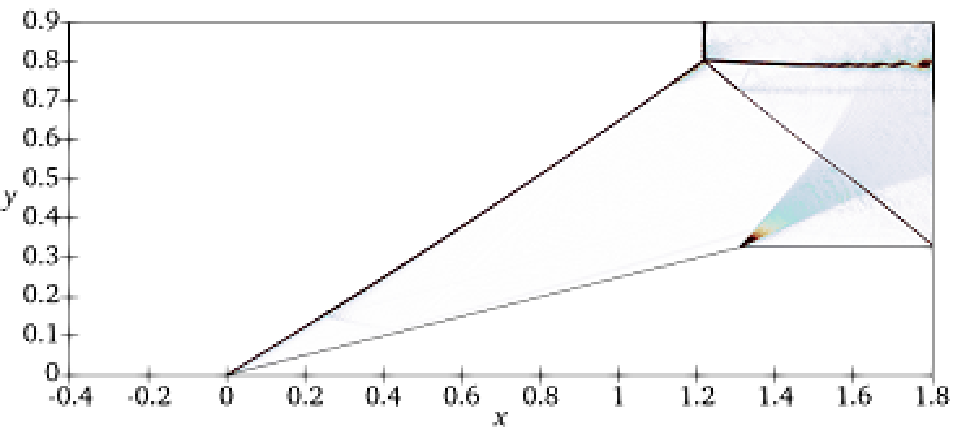}
        \caption{$\theta=14^{\circ}$}\label{fig:fig12_14}
	\end{subfigure}
	\begin{subfigure}{0.27\linewidth}
	    \centering
        \includegraphics[width=0.99\linewidth]{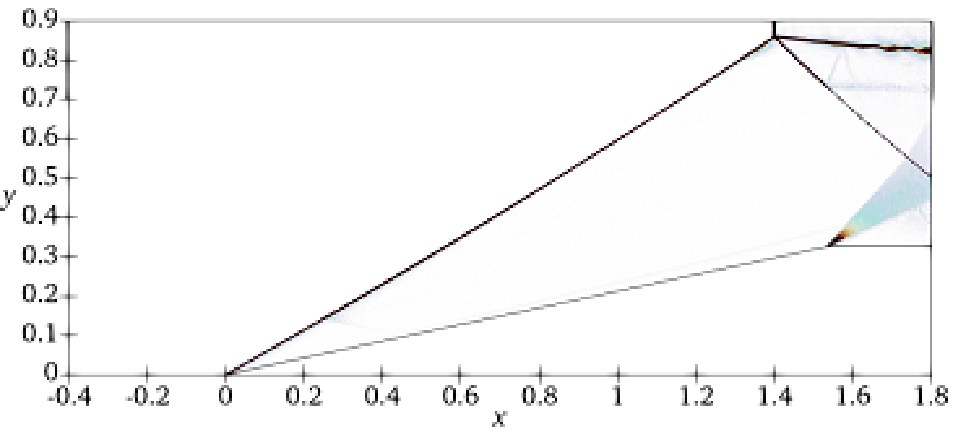}
        \caption{$\theta=12^{\circ}$}\label{fig:12_12}
	\end{subfigure}
	\begin{subfigure}{0.27\linewidth}
	    \centering
        \includegraphics[width=0.99\linewidth]{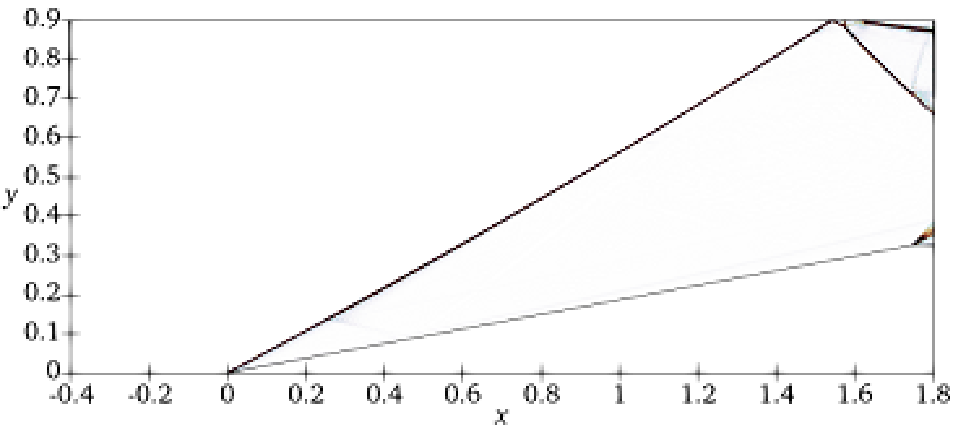}
        \caption{$\theta=10.57^{\circ}$}\label{fig:fig12_10_57}
	\end{subfigure}
	\caption{Velocity gradient contours of the dynamic shock system at a relatively high non-dimensional angular velocity, $\kappa = 1.0$. }\label{fig:fig12}
\end{figure}

\begin{figure}[H]
	\centering
	\begin{subfigure}{0.27\linewidth}
	    \centering
        \includegraphics[width=0.99\linewidth]{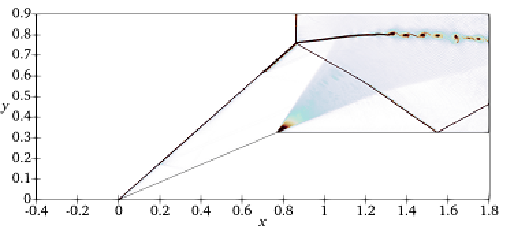}
        \caption{$\theta=23^{\circ}$}\label{fig:fig10_23}
	\end{subfigure}
	\begin{subfigure}{0.27\linewidth}
	    \centering
        \includegraphics[width=0.99\linewidth]{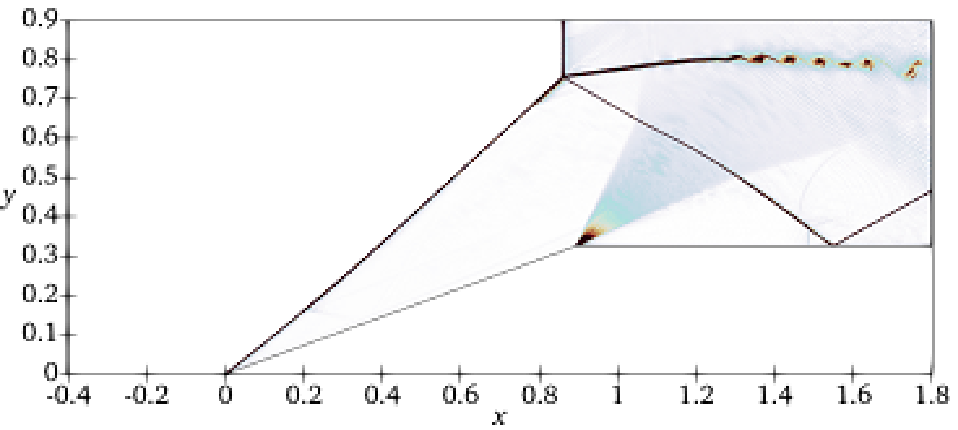}
        \caption{$\theta=20^{\circ}$}\label{fig:fig10_20}
	\end{subfigure}
	\begin{subfigure}{0.27\linewidth}
    	\centering  
        \includegraphics[width=0.99\linewidth]{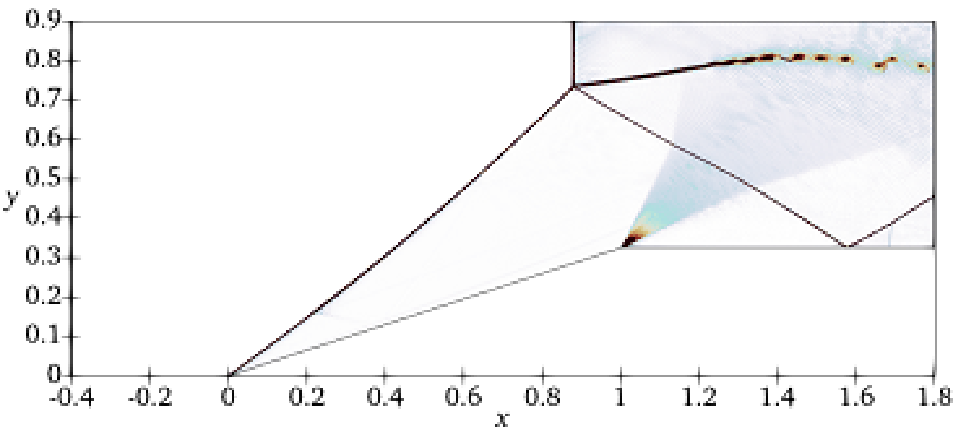}
        \caption{$\theta=18^{\circ}$}\label{fig:fig10_18}
	\end{subfigure}\\
	\begin{subfigure}{0.27\linewidth}
	    \centering
        \includegraphics[width=0.99\linewidth]{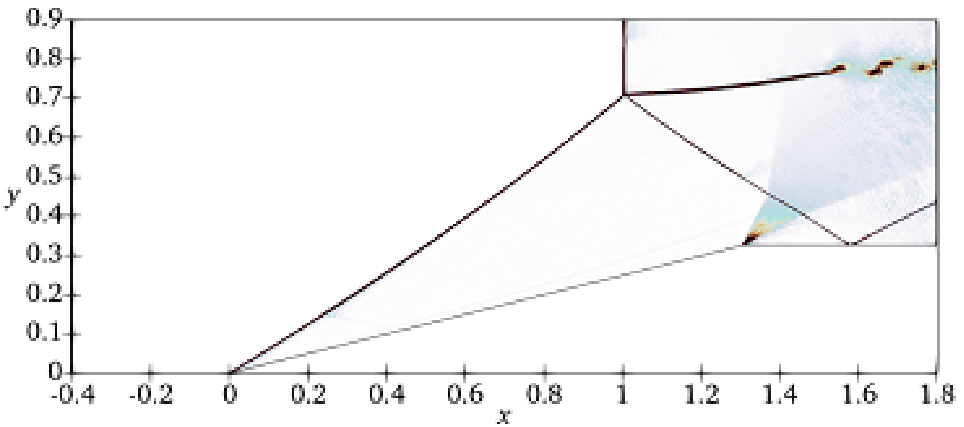}
        \caption{$\theta=14^{\circ}$}\label{fig:fig10_14}
	\end{subfigure}
	\begin{subfigure}{0.27\linewidth}
	    \centering
        \includegraphics[width=0.99\linewidth]{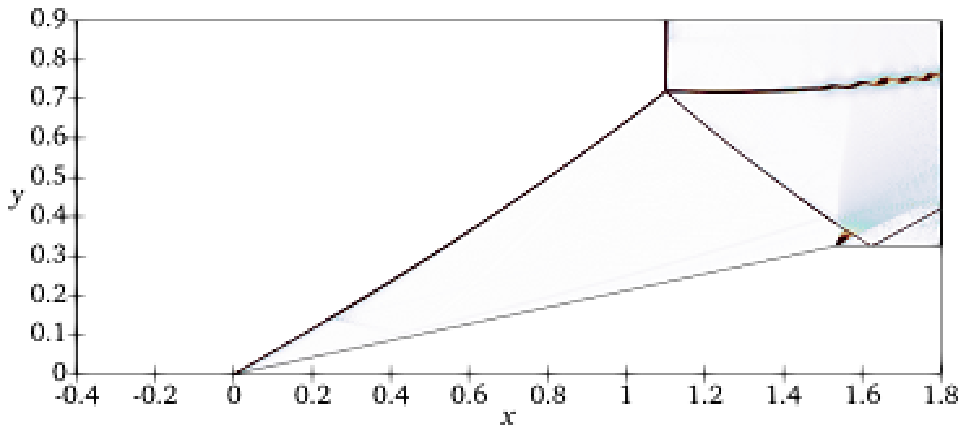}
        \caption{$\theta=12^{\circ}$}\label{fig:fig10_12}
	\end{subfigure}
	\begin{subfigure}{0.27\linewidth}
	    \centering
        \includegraphics[width=0.99\linewidth]{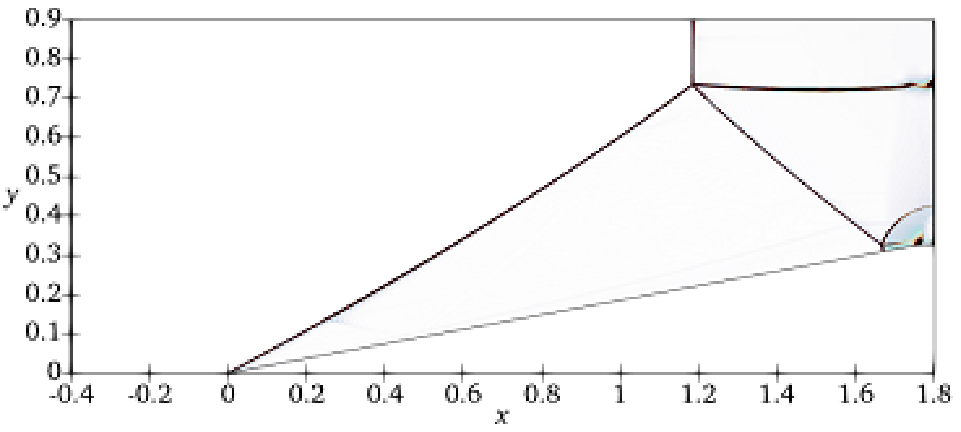}
        \caption{$\theta=10.5^{\circ}$}\label{fig:fig10_10_5}
	\end{subfigure}
	\caption{Velocity gradient contours of the dynamic shock system at a high non-dimensional angular velocity, $\kappa = 2.0$. }\label{fig:fig10}
\end{figure}

Analytically, when a sharp compression ramp opposes a supersonic flow, there are many possible shock wave systems, including the RR and the MR as summarized by Mouton \cite{mouton2007transition}. The wedge angle and the free-stream Mach number influence the formation of a certain configuration. Figure \ref{fig:fig9} shows a comparison between the dynamic von-Neumann transition wave angle using different motion speeds to the theoretical transition criteria on the Duel Solution Domain, DSD. As $\theta$ decreased with a certain value of $\kappa$, the wave angle decreased, placing the dynamic transition tangent wave angle, $\beta_{t_{tang}}$, beyond the theoretical von-Neumann condition. For low motion rates, such as $\kappa=0.1$, the $\beta_{t_{tang}}$ was just below the analytical limit. Using higher non-dimensional angular velocities, such as $\kappa=1.0$, resulted in increasing the gap between the transition wave angle and the physical steady-state limit. This was because of the dynamic shock system's extreme lag.

\section{Conclusion}
The aim of the current paper is to numerically study the dynamic transition from MR to RR for an inviscid supersonic flow of $M_{\infty}=3$ over a moving wedge. The motion was achieved, by decreasing the wedge angle with different non-dimensional angular velocities, $\kappa$, and keeping the wedge height fixed. The dynamic transitional angles and the dynamic hysteresis were analyzed to investigate the phenomenon. This research work concludes that the non-dimensional angular velocity of the motion lagged the shock system, causing the occurrence of the transitional angle below the theoretical von-Neumann criterion in the Dual-Solution Domain. Moreover, the strength of the lag is clearly observed in the Mach stem height, where the transition did not happen at $\kappa=1.5$ and $2.0$ with our current geometric limits. Further, the Sonic and von-Neumann criteria happened very close to each other as the difference in the transitional angles was within a degree.
\bibliography{Margha}
\end{multicols*}


\end{document}